An Analysis of Universality in Blackbody Radiation
Pierre-Marie Robitaille, Ph.D.*
Chemical Physics Program
The Ohio State University, Columbus, Ohio 43210

*Through the formulation of his law of thermal emission, Kirchhoff conferred upon blackbody radiation the quality of universality [G. Kirchhoff, Annalen der Physik **109**, 275 (1860)]. Consequently, modern physics holds that such radiation is independent of the nature and shape of the emitting object. Recently, Kirchhoff's experimental work and theoretical conclusions have been reconsidered [P.M.L. Robitaille. IEEE Transactions on Plasma Science. **31(6)**, 1263 (2003)]. In this work, Einstein's derivation of the Planckian relation is reexamined. It is demonstrated that claims of universality in blackbody radiation are invalid.*

From the onset, blackbody radiation was unique in possessing the virtue of universality [1,2]. The nature of the emitting object was irrelevant to emission. Planck [3], as a student of Kirchhoff, adopted and promoted this concept [4,5]. Nonetheless, he warned that objects sustaining convection currents should not be treated as blackbodies [5].

As previously discussed in detail [6], when Kirchhoff formulated his law of thermal emission [1,2], he utilized two extremes: the perfect absorber and the perfect reflector. He had initially observed that all materials in his laboratory displayed distinct emission spectra. Generally, these were not blackbody in appearance and were not simply related to temperature changes. Graphite, however, was an anomaly, both for the smoothness of its spectrum and for its ability to simply disclose its temperature. Eventually, graphite's behavior became the basis of the laws of Stefan [7], Wien [8] and Planck [3].

For completeness, the experimental basis for universality is recalled [1,2,5,6]. Kirchhoff first set forth to manufacture a box from graphite plates. This enclosure was a near perfect absorber of light ($\varepsilon = 1$, $\kappa = 1$). The box had a small hole, through which radiation escaped. Kirchhoff placed various objects in this device. The box would act as a transformer of light [6]. From the graphitic light emitted, Kirchhoff was able to gather the temperature of the enclosed object once thermal equilibrium had been achieved. A powerful device had been constructed to ascertain the temperature of any object. However, this scenario was strictly dependent on the use of graphite.

Kirchhoff then sought to extend his findings [1,2,5]. He constructed a second box from metal, but this time the enclosure had perfectly reflecting walls ($\varepsilon = 0$, $\kappa = 0$). Under this second scenario, Kirchhoff was never able to reproduce the results he had obtained with the graphite box. No matter how long he waited, the emitted spectrum was always dominated by the object enclosed in the metallic box. The second condition was unable to produce the desired spectrum.

As a result, Kirchhoff resorted to inserting a small piece of graphite into the perfectly reflecting enclosure [5]. Once the graphite particle was added, the spectrum changed to that of the classic blackbody. Kirchhoff believed he had achieved universality. Both he, and later, Planck, viewed the piece of graphite as a "catalyst" which acted only to increase the speed at which equilibrium was achieved [5]. If only time was being compressed, it would be mathematically appropriate to remove the graphite particle and to assume that the perfect reflector was indeed a valid condition for the generation of blackbody radiation.

However, given the nature of graphite, it is clear that the graphite particle was in fact acting as a perfect absorber. Universality was based on the validity of the experiment with the perfect reflector, yet, in retrospect, and given a modern day understanding of catalysis and of the speed of light, the position that the graphite particle acted as a catalyst is untenable. In fact, by adding a perfect absorber to his perfectly reflecting box, it was as if Kirchhoff lined the entire box with graphite. He had unknowingly returned to the first case. Consequently, universality remains without any experimental basis.

Nonetheless, physics has long since dismissed the importance of Kirchhoff's work [9]. The basis for universality, no longer rests on the experimental proof [i.e. 9], but rather on Einstein's theoretical formulation of the Planckian relation [10, 11]. It has been held [i.e. 9] that with Einstein's derivation, universality was established beyond doubt based strictly on a theoretical platform. Consequently, there appears to no longer be any use for the experimental proof formulated by Kirchhoff [1,2,5]. Physics has argued [9] that Einstein's derivation of the Planckian equations had moved the community beyond the limited confines of Kirchhoff's enclosure. Einstein's derivation, at least on the surface, appeared totally independent of the nature of the emitting compound. Blackbody radiation was finally free of the constraints of enclosure.

In his derivation of the Planckian relation, Einstein has recourse to his well-known coefficients [10,11]. Thermal equilibrium and the quantized nature of light (E=hυ ) are also used. All that is required appears to be 1) transitions within two states, 2) absorption, 3) spontaneous emission, and 4) stimulated emission. However, Einstein also requires that gaseous atoms act as perfect absorbers and emitters or radiation. In practice, of course, isolated atoms can never act in this manner. In all laboratories, isolated groups of atoms act to absorb and emit radiation in narrow bands and this only if they possess a dipole moment. This is well-established in the study of gaseous emissions [12]. As such, Einstein's requirement for a perfectly absorbing atom, knows no physical analogue on earth. In fact, the only perfectly absorbing materials known, exist in the condensed state. Nonetheless, for the sake of theoretical discussion, Einstein's perfectly absorbing atoms could be permitted.

In his derivation, Einstein also invokes the requirement of thermal equilibrium with a Wien radiation field [8], which of course, required enclosure [1,2]. However, such a field is uniquely the product of the solid state. To be even more specific, a Wien's radiation field is currently produced with blackbodies typically made either from graphite itself or from objects lined with soot. In fact, it is interesting that graphite (or soot) maintain a prominent role in the creation of blackbodies currently used at the National Bureau of Standards [13-17].

Consequently, through his inclusion of a Wien's radiation field [8], Einstein has recourse to a physical phenomenon which is known to be created exclusively by a solid. Furthermore, a Wien's field, directly involves Kirchhoff's enclosure. As a result, claims of universality can no longer be supported on the basis of Einstein's derivation of the Planckian relation. A solid is required. Therefore, blackbody radiation remains exclusively a property of the solid state. The application of the laws of Planck [3], Stefan [7] and Wien [8] to non-solids is without both experimental and theoretical justification.

*130 Means Hall, 1654 Upham Drive, The Ohio State University, Columbus, Ohio 43210
robitaille.1@osu.edu, 1-614-293-3046